\documentclass[prd,preprint,superscriptaddress,preprintnumbers,eqsecnum,showpacs,nofootinbib,nobibnotes]{revtex4-1}
\usepackage{amsfonts,amsmath,amssymb,bm}
\usepackage[table]{xcolor}
\usepackage{graphicx} 
\usepackage{mathtools}
\usepackage{boondox-cal}
\usepackage{slashed}
\usepackage{xcolor}

\usepackage[utf8]{inputenc}
\usepackage{slashed}

\newcommand{\be}{\begin{equation}}
\newcommand{\bea}{\begin{eqnarray}}
\newcommand{\ba}{\begin{align}}
\newcommand{\ee}{\end{equation}}
\newcommand{\eea}{\end{eqnarray}}
\newcommand{\ea}{\end{align}}

\definecolor{zero2}{rgb}{0.88,0.88,.88}

\def\1eq#1{Eq.~(\ref{#1})}

\def\2eqs#1#2{Eqs.~(\ref{#1}) and~(\ref{#2})}
\def\3eqs#1#2#3{Eqs.~(\ref{#1}),~(\ref{#2}) and~(\ref{#3})}
\def\4eqs#1#2#3#4{Eqs.~(\ref{#1}),~(\ref{#2}),~(\ref{#3}) and~(\ref{#4})}

\def\s#1{{\scriptscriptstyle #1}}

\def\G{\Gamma}

\def\s{\mathcal{s}}

\def\hphi0{{\hat\phi}_0}

\makeatletter
\def\user@resume{resume}
\def\user@intermezzo{intermezzo}
\newcounter{previousequation}
\newcounter{lastsubequation}
\newcounter{savedparentequation}
\def\CT@@do@color{%
	\global\let\CT@do@color\relax
		\@tempdima\wd\z@
		\advance\@tempdima\@tempdimb
		\advance\@tempdima\@tempdimc
		\advance\@tempdimb\tabcolsep
		\advance\@tempdimc\tabcolsep
		\advance\@tempdima1.5\tabcolsep
	\kern-1.5\@tempdimb
	\leaders\vrule
	\hskip\@tempdima\@plus  1fill
	\kern-1.5\@tempdimc
	\hskip-\wd\z@ \@plus -1fill }
\makeatother

\begin{document}

\title{
A New Stability Equation for the 
Abelian Higgs-Kibble Model\\
with a Dim.6 Derivative Operator
}

\date{February 19, 2023}

\author{A. Quadri}
\email{andrea.quadri@mi.infn.it}
\affiliation{INFN, Sezione di Milano, via Celoria 16, I-20133 Milano, Italy}

\begin{abstract}
\noindent
We show that the 
dynamics of the scalar Higgs field in the Abelian Higgs-Kibble model, supplemented with a dimension 6
derivative operator, can be 
constrained at the quantum level
by a certain stability equation.
The latter holds true in the Landau
gauge and is derived within the
recently proposed extended field formalism where the physical scalar is described by a gauge-invariant field variable.
Physical implications of the stability equation are  discussed.\\
\\
Submitted to the Proceedings of the A.A.Slavnov
Memorial Conference, December 2022.
\\

In honour of Academician Professor A.A.Slavnov
\end{abstract}


\maketitle

\section{Introduction}

The question of whether one can give sense to a non power-counting
renormalizable theory is an old one that still deserves some investigation, due to both phenomenological and theoretical reasons.

On the one hand, 
the forthcoming LHC experimental program will allow us for the first time to address the measurement of the Higgs self-couplings~\cite{Cepeda:2019klc}, thus testing experimentally the predictions of the Standard Model (SM) in the Higgs self-interaction sector. 

As is well known, the SM prediction
about the quartic Higgs potential is a consequence of the combined requirement
of gauge invariance and  power-counting renormalizability, that limits the admissible terms in the classical action to be gauge-invarant operators of dimension~$\leq 4$.

Gauge invariance alone is in fact compatible
with a whole set of additional higher dimensional
Beyond the Standard Model (BSM)
interactions, that have been widely
studied in recent years in the context of the SM effective field theories~\cite{Buchmuller:1985jz,Alonso:2013hga,Brivio:2017vri}.

The fundamental difference between a power-counting renormalizable model like the SM and effective field theories, involving higher dimensional operators suppressed by some large energy scale $\Lambda$, concerns the consistent subtraction of UV divergences once radiative corrections are taken into account.

Under very general assumptions, it is known~\cite{Gomis:1995jp} that
effective field theory models can be recursively made finite, order by order in the loop expansion, by
adding counter-terms that are local formal power series in the fields, the external sources and their derivatives. These counter-terms satisfy the defining Batalin-Vilkovisky master equation~\cite{Gomis:1994he}, or equivalently the Slavnov-Taylor identity~\cite{Slavnov:1972fg,Taylor:1971ff}, generalizing at the quantum level the classical Becchi-Rouet-Stora-Tyutin (BRST) symmetry.
The counter-terms can be  subdivided into two classes: the generalized (i.e. non-linear and in general also non-polynomial) field redefinitions and  the renormalizations of the gauge couplings of the (infinite) number of gauge-invariant operators
of higher and higher dimension,
that one must add in order to remove the UV divergences of increasing dimension 
as more and more loops are included.

This is at variance with the power-counting renormalizable models, for which the linear renormalizations of the fields and of the couplings
 of the operators of dim.$\leq 4$
 are sufficient in order to make the theory finite, order by order in the loop expansion, while respecting all the defining symmetries of the model.
 
 On the other hand, from a theoretical point of view deciding whether the number of independent UV divergent amplitudes is finite or not can be a tricky task.
 In fact the appropriate choice of field variables can reveal hidden relations between those amplitudes.
 The simplest illustrative case is the massless scalar free field described by the action
 $$ S = \int d^4 x \, \frac{1}{2}
 \partial^\mu \phi \partial_\mu \phi \, .$$
 A generalized invertible field redefintion of the form
\begin{align}
\phi \rightarrow h(\phi) \phi \, ,
\label{f.redef}
\end{align}
with $h$ an arbitrary polynomial in $\phi$ such that 
$h(0)=1$, leads to a {\it prima facie} non renormalizable theory,
namely
$$ S = \int d^4x \, \Big [ 
\frac{h^2}{2} + 
\phi \dot{h} 
+ \frac{\phi^2}{2} \dot{h}^2
\Big ] \partial^\mu \phi \partial_\mu \phi \, , $$
where the dot denotes differentiation with respect to $\phi$.

Not unexpectedly, provided that the change of variables (\ref{f.redef}) is consistently implemented in the path-integral via the so-called Equivalence Theorem~~\cite{Kallosh:1972ap,Kamefuchi:1961sb,Ferrari:2002kz}, the second action give rise to a quantized theory that is physically equivalent to the 
original free field theory.

Less trivially, the use of a gauge-invariant dynamical field to describe the physical Higgs mode~\cite{Frohlich:1980gj,Frohlich:1981yi,Binosi:2017ubk,Binosi:2019olm,Binosi:2019nwz,Dudal:2019pyg,Dudal:2020uwb,Binosi:2020unh,Dudal:2021pvw,Dudal:2021dec},
at variance with the usual description through a non-gauge invariant variable (respectively a complex
scalar field for the Abelian case,
the SU(2) Higgs doublet for the SM)
has recently led to the so-called extended formalism~\cite{Binosi:2017ubk,Binosi:2019olm,Binosi:2019nwz,Binosi:2020unh}, in which a number of additional relations between amplitudes are captured by a powerful differential equation that controls the whole dependence of the theory on the coupling $z$ of the dim.6 derivative operator $\phi^\dagger \phi \square \phi^\dagger \phi$.

For that purpose one introduces a dynamical field $X_2$
that is on-shell equivalent to the gauge-invariant combination
\begin{align}
\phi^\dagger \phi - \frac{v^2}{2} \sim v X_2 \, , 
\end{align}
with $v$ being the vacuum expectation value of $\phi$. The Lagrange multiplier field
enforcing this condition is denoted by $X_1$.

The key point is that one can then recast the whole dependence on the 
coupling $z$ into a quadratic contribution in $X_2$, namely
\begin{align}
    -\frac{z}{2v^2} \phi^\dagger \phi \square \phi^\dagger \phi \sim -\frac{z}{2} X_2 \square X_2 \, ,
    \label{d6.op}
\end{align}
so that $z$ will never appear in the interaction terms but only via the $X_2$-propagator.

This is major combinatorial simplification, that allows one to
write a differential equation for the amplitudes fixing their $z$-dependence once the amplitudes
of the (power-counting renormalizable) model at $z=0$ are
known.

Therefore the amplitudes of the
non-power-counting renormalizable theory at $z\neq 0$ are in fact
uniquely predicted in terms of a finite number of physical parameters, namely those of  the power-counting renormalizable theory and $z$.
As such, the model at $z\neq 0$
can be deemed renormalizable despite
not being power-counting renormalizable.

In the present paper we take 
a step further into the investigation of the properties of the Abelian theory
at $z\neq 0$, as a playground towards the study of the non-Abelian case and the
electroweak theory.

We will show that an additional stability equation exists in the Landau gauge, controlling the dependence of the vertex functional on the 
field $\phi$ (in addition to the
$X_{1,2}$-equations of motion).

The stability equation selects a vanishing value of the parameter $m^2$, that governs the quartic operator $\Big ( \phi^\dagger \phi - \frac{v^2}{2} \Big )^2$. The latter is generated
off-shell in the extended formalism and one can prove that  physical 
gauge-invariant observables are perturbatively $m^2$-independent once the equations of motion for $X_{1,2}$ are imposed.

The condition $m=0$ in turn ensures that the classical 
scalar potential is semi-positive defined along the directions spanned by the auxiliary fields in the extended formalism. This property is not violated by quantum corrections, as 
a consequence of the definining functional identities of the theory.

The paper is organized as follows.
Sect.~\ref{sec.intro} is devoted to a concise presentation of the Abelian Higgs-Kibble model in the
extended formalism in the presence of the dim.6 operator $\phi^\dagger \phi \square \phi^\dagger \phi$.
In Sect.~\ref{sec:stability}
we derive the stability equation for $\phi$.
In Sect.~\ref{sec:zdiffeq} the $z$-differential equation is presented.
In Sect.~\ref{sec:potential} we comment on the stability of the classical potential in the extended field space and show that the latter is preserved at the quantum level
for vanishing $m^2$.
Conclusions are finally given in Sect.~\ref{sec:conclusions}.

\section{Classical Action}\label{sec.intro}

We consider the Abelian Higgs-Kibble (HK) model~\cite{Becchi:1974md,Becchi:1974xu} extended  with the dimension 6 operator in Eq.(\ref{d6.op}).
We will use a gauge-invariant coordinate in order to describe the physical scalar mode, in accord with the formalism of~\cite{Quadri:2006hr,Quadri:2016wwl,Binosi:2017ubk}.

Moreover we will introduce a set of auxiliary complex vector fields, $K_\mu$ and $K_\mu^\dagger$.
The advantage of this representation is that one can obtain
a stability equation for the complex Higgs field 
$\phi$, as we will show in Sect.~\ref{sec:stability}.

In components
$\phi = \frac{1}{\sqrt{2}} (\sigma + v + i \chi)$, $\chi$ being the pseudo-Goldstone boson and $v$ the vacuum expectation value (v.e.v), so that the real field $\sigma$ has vanishing v.e.v.

In order to use the gauge-invariant field $X_2$ 
one also needs the  Lagrange multiplier $X_1$ to obtain the action
\begin{align}
	\G^{(0)}  = 
	   \int \!\mathrm{d}^4x \, &\Bigg [ -\frac{1}{4} F^{\mu\nu} F_{\mu\nu} - K_\mu^\dagger K^\mu + K^\dagger_\mu D^\mu \phi + (D_\mu \phi)^\dagger K^\mu
    - \frac{M^2-m^2}{2} X_2^2 - \frac{m^2}{2v^2} \left ( \phi^\dagger \phi - \frac{v^2}{2} \right )^2 \nonumber \\
	&  +\frac{z}{2} \partial^\mu X_2 \partial_\mu X_2 - \bar c (\square + m^2) c + \frac{1}{v} (X_1 + X_2) (\square + m^2) \left ( \phi^\dagger \phi - \frac{v^2}{2} - v X_2 \right ) \nonumber \\
	& + \frac{\xi b^2}{2} -  b \left ( \partial A + \xi e v \chi \right ) + \bar{\omega}\left ( \square \omega + \xi e^2 v (\sigma + v) \omega\right ) \nonumber \\
    & + \phi^{*} ~i e \omega \phi + \phi^{\dagger *} (-i e \omega \phi^\dagger) 
    + K^{\mu *} ~i e \omega K_\mu + K^{\mu \dagger*} (-i e \omega K_\mu^\dagger) \nonumber \\
    & + \bar c^* \Big ( \phi^\dagger \phi - \frac{v^2}{2} - v X_2 \Big )
 \Bigg ].
	\label{tree.level}
\end{align}
In the above equation $D_\mu$ is the covariant derivative
\begin{align}
	D_\mu = \partial_\mu - i e A_\mu .
\end{align} 
$A_\mu$ is the Abelian gauge connection, $F_{\mu\nu} = \partial_\mu A_\nu - \partial_\nu A_\mu$ is the field strength. $K^\mu, K^{\mu \dagger}$ transform as $\phi$ under the $U(1)$ gauge transformation.

Notice that  after the shift $K_\mu \rightarrow K_\mu'\equiv K_\mu - D_\mu \phi$ the redefined vectors $K'_\mu, {K_\mu'}^\dagger$
have constant propagators, so no new dynamical degrees of freedom
are introduced in the theory.

The third line in Eq.(\ref{tree.level}) contains the gauge-fixing and the usual ghost-antighost terms. The corresponding BRST symmetry is described in Sect.~\ref{sec:brst}.
Finally the fourth line contains the external sources coupled to the non-linear BRST
transformation of the quantized fields (the so-called antifields~\cite{Gomis:1994he}), while the last one contains the 
external source $\bar c^*$ that is required in order to formualte at the quantum level
the equations of motion for the fields $X_{1,2}$.

$X_1$ enforces on-shell the condition
\begin{align} 
X_2 \sim \frac{1}{v} \left ( \phi^\dagger \phi - \frac{v^2}{2} \right ).
\label{os.constraint}
\end{align}
Consequently, $X_2$ can be thought of as a field coordinate parametrizing  the scalar gauge-invariant combination $\phi^\dagger \phi - \frac{v^2}{2}$; in particular, at the linearized level, $X_2 \sim \sigma$.
We notice that going on-shell with $X_1$ yields a Klein-Gordon equation
$$ (\square + m^2) \left ( \phi^\dagger \phi - \frac{v^2}{2} - v X_2 \right ) = 0, $$
whose most general solution is $X_2 = \frac{1}{v} \left ( \phi^\dagger \phi - \frac{v^2}{2} \right ) + \eta$, $\eta$ being a scalar field of mass $m$.
However in
perturbation theory one can show that the correlators of the mode 
$\eta$ with any gauge-invariant operators vanish~\cite{Binosi:2019olm}, so that
one can safely set $\eta= 0$.

%
%
%

By eliminating in \1eq{tree.level} both $X_1, X_2$ 
and $K^\mu, K^{\mu \dagger}$ via their equations of motion one recovers the usual vertex functional of the Abelian Higgs-Kibble model with the dimension 6 derivative operator 
\begin{align}
    \frac{z}{2} \partial^\mu X_2 \partial_\mu X_2 \sim
    \frac{1}{2 v^2} \partial^\mu  \left ( \phi^\dagger \phi - \frac{v^2}{2} \right ) \partial_\mu  \left ( \phi^\dagger \phi - \frac{v^2}{2} \right ) \, .
\end{align}
In particular, the two mass parameters $m$ and $M$ in the first line of~\1eq{tree.level} are chosen in such a way that by going on-shell with the Lagrange multiplier $X_1$ one recovers the usual quartic Higgs potential
$ - \frac{M^2}{2v^2} 
 \left ( \phi^\dagger \phi - \frac{v^2}{2} \right )^2 \, .$
The only physical parameter is thus $M$. In fact, it can be checked that the correlators of physical observables after going on-shell with the fields $X_{1,2}$ do not depend on $m$~\cite{Binosi:2019olm}. 

\subsection{BRST symmetry}\label{sec:brst}

The tree-level vertex functional in Eq.(\ref{tree.level}) is invariant under two distinct BRST symmetries. The first one
is associated with gauge U(1) invariance and reads
\begin{align}
    s A_\mu = \partial_\mu \omega ;  \quad s\phi = i e \omega \phi  ;  
    \quad s K_\mu = i e \omega K_\mu  ; 
    \quad s \bar \omega = b ;  \quad s b = 0,
    \label{brst.gauge}
\end{align}
all other fields being invariant.
The corresponding  ghost is denoted by $\omega$.

The second one is a consequence of the constraint U(1) BRST symmetry 
\begin{align}
    &\s X_1 = v c, & &\s c = 0, &
    \s \bar c = \frac{1}{v}
    \left ( \phi^\dagger \phi - \frac{v^ 2}{2} - v X_2 \right ),
    \label{brst.constr}
\end{align}
all other fields and external sources being invariant under $\s$.
$c, \bar c$ are the constraint U(1) ghost and antighost fields.
The constraint U(1) BRST symmetry ensures that the extension of the scalar sector via the fields $X_{1,2}$ does not introduce additional physical degrees of freedom~\cite{Binosi:2022ycu}.  

The ghost number is assigned as follows: $\omega, c$ have ghost number $1$, $\bar c, \bar \omega$ have ghost number $-1$.
The antifields $\phi^*, \phi^{\dagger *}, K^\mu, K^{\mu\dagger *}$
have ghost number $-1$. All the other fields and external sources
have ghost number zero.

It is convenient to combine both the gauge and the constraint U(1) BRST
symmetry into a single BRST differential $\tilde{s}$, given by
$$\tilde{s} = \s + s \, .$$
$\tilde{s}$ is nilpotent as a consequence of the fact that 
$\s, s$ anti-commute.

Invariance under $\tilde{s}$ leads to the following
Slavnov-Taylor identity:
\begin{align}
 {\cal S}(\G)  = \int \mathrm{d}^4x \, \Big [ 
&	\partial_\mu \omega \frac{\delta \G}{\delta A_\mu} 
 + \frac{\delta \G}{\delta \phi^{*\dagger}} \frac{\delta \G}{\delta \phi^\dagger}  
 + \frac{\delta \G}{\delta \phi^*} \frac{\delta \G}{\delta \phi}
 + \frac{\delta \G}{\delta K^{\mu *\dagger}} \frac{\delta \G}{\delta K_\mu^\dagger}  
 + \frac{\delta \G}{\delta K^{\mu*}} \frac{\delta \G}{\delta K_\mu}
	+ b \frac{\delta \G}{\delta \bar \omega} \nonumber \\
 &  +
 v c \frac{\delta \G}{\delta X_1} 
 + \frac{\delta \G}{\delta \bar c^*}\frac{\delta \G}{\delta \bar c} 
 \Big ] = 0 \, .
\label{sti} 
\end{align}
In addition to the ST identity, the gauge-fixing sector is 
controlled by the following stability equations:
\begin{itemize}
\item The $b$-equation:
\begin{eqnarray}
	\frac{\delta \G}{\delta b} =\xi b - \partial A -  \xi e v \chi;
	\label{b.eq}
\end{eqnarray}
\item The antighost equation:
\begin{eqnarray}
	\frac{\delta \G}{\delta \bar \omega} = \square \omega +  i \xi \frac{ev}{\sqrt{2}} \Big ( 
        \frac{\delta \G}{\delta \phi^{\dagger *}} - \frac{\delta \G}{\delta \phi^*} \Big ) \, .
	\label{antigh.eq}
\end{eqnarray}   
In the above equation we have used the fact that $\chi = \frac{i}{\sqrt{2}} (\phi^\dagger - \phi)$ in order to re-express the BRST transformation $s \chi$ in terms
of the BRST transformations of $\phi, \phi^\dagger$ and hence
in terms of the functional derivatives
w.r.t. $\phi^*,\phi^{\dagger *}$.
\end{itemize}
\section{Stability equation for the $\phi$-field}\label{sec:stability}
The introduction of the complex vector field $K_\mu$ allows one
to derive a further stability equation for the field $\phi$,
in the particular case where the auxiliary parameter $m$ vanishes.
We work in the Landau gauge $\xi=0$.

For $m=0$ the dependence on $\phi$ of the classical vertex functional 
in Eq.(\ref{tree.level})  is linear
with the exception of the last term in the second line, giving rise
to a quadratic dependence.

It turns out that it is possible to derive a stability equation for $\phi$
that holds true at the quantum level, provided that a suitable set of 
external sources is introduced in order to handle the renormalization
of the composite operators controlling such a quadratic dependence on $\phi$.

For that purpose we take a functional derivative of Eq.(\ref{tree.level}) and
find (notice that in Landau gauge there is no contribution from the gauge-fixing term)
\begin{align}
    \frac{\delta \G^{(0)}}{\delta \phi} = - (D^\mu K_\mu)^\dagger + i e \omega \phi^* + \bar c^* \phi^\dagger + \frac{1}{v} \square (X_1 + X_2) \phi^\dagger \, .
    \label{stability.eq}
\end{align}
The terms in the r.h.s. of the above equation 
depending on the anti-fields $\phi^*, \bar c^*$
 are linear
 in the quantum fields and thus they do not receive radiative corrections.

On the contrary, the first and last term are quadratic in the fields
 and therefore we need to couple them in the classical action to 
 suitable external sources in order to control their
 renormalization.
 
 Let us consider first the operator $\square (X_1 + X_2) \phi^\dagger$. We couple it to the complex external source $N$.
In order to preserve the ST identity, the source $N$ must be
 paired into a BRST doublet~\cite{Gomis:1994he,Quadri:2002nh} to its BRST partner $L$ with ghost number $-1$:
 \begin{align}
     \tilde{s} L = N \, , \quad \tilde{s} N = 0 \, .
 \end{align}
Hence we can add to the classical action the term
\begin{align}
    \int d^4x \, & \tilde{s} \Big [L ~\frac{1}{v} \square (X_1 + X_2)  \phi^\dagger \Big ]  + h.c.  \nonumber \\
 &
    = \int d^4x \, \Bigg [ \frac{1}{v} N~ \square (X_1 + X_2)  \phi^\dagger - L ~ \square c ~ \phi^\dagger + \frac{1}{v} L  
    \square (X_1 + X_2)~ ie \omega \phi^\dagger + h.c.\Bigg ] \, .
    \label{add.1}
\end{align}
By taking the functional derivative of Eq.(\ref{add.1})
w.r.t. $\phi$  a new operator arises, namely
\begin{align}
    -\frac{i e}{v} L^\dagger \square (X_1 + X_2) \omega \, .
\end{align}
Again we need to define this operator by coupling it into the classical action to the source $P$, to be paired into a BRST
doublet with its partner $Q$ in order to respect the ST identity:
\begin{align}
    \tilde{s} Q = P \, , \quad \tilde{s} P = 0 \, .
\end{align}
$Q$ has ghost number $-2$, while $P$ has ghost number $-1$.

The new set of external sources is then finally given by
\begin{align}
    \int d^4x \, & \Bigg \{ 
    \tilde{s}  ( Q \square (X_1 + X_2) \omega ) +
    \Big [ 
    \tilde{s} \Big ( L ~\frac{1}{v} \square (X_1 + X_2)  \phi^\dagger \Big )  + h.c. \Big ] \Bigg \}  \nonumber \\
 &
    = \int d^4x \, \Bigg [ 
    P \square (X_1+ X_2) \omega + Q \square ( v c) \omega 
    \nonumber \\ 
    & \qquad \qquad + 
    \frac{1}{v} N~ \square (X_1 + X_2)  \phi^\dagger - L ~ \square c ~ \phi^\dagger + \frac{1}{v} L  
    \square (X_1 + X_2)~ ie \omega \phi^\dagger + h.c.\Bigg ] \, .
    \label{add.2}
\end{align}

The treatment of the operator $D^\mu K_\mu$ is simpler, because just one BRST doublet of external sources
\begin{align}
    \tilde{s} R = T \, , \qquad \tilde{s} T = 0 
\end{align}
is required. The
sources $R,T$ contribute to the classical action 
via the term
\begin{align}
    \int d^4 x \, \tilde s \Big [  (D^\mu K_\mu)^\dagger R + h.c. \Big ] =
    \int d^4 x \, \Big [ (D^\mu K_\mu)^\dagger T - i e \omega (D^\mu K_\mu)^\dagger R + h.c. \Big ] \, .
\end{align}
The algebra of operators eventually closes  and we can write down the
complete stability equation for $\phi$:
\begin{align}
    \frac{\delta \G}{\delta \phi} = 
    \frac{\delta \G}{\delta N} - \frac{\delta \G}{\delta T} - \frac{ie}{v} L^\dagger\frac{\delta \G}{\delta P} 
     + i e \omega \phi^* + \bar c^* \phi^\dagger - L^\dagger \square c \, 
    \label{stability.eq.full}
\end{align}
and a corresponding conjugate relation for the $\phi^\dagger$-stability equation.

Notice that no additional sources are needed in order to control operators that are linear in the quantized fields in the r.h.s. of Eq.(\ref{stability.eq.full}) and thus no new source is required to define
them.

A comment is in order here. One can extend the construction also to a general $R_\xi$-gauge.
However an additional source coupled to the ghost condensate opertor $\bar \omega \omega$ is required
in order to formulate the stability equation and in very much the same way as discussed above
further terms are generated that modify in particular the Nakanishi-Lautrup $b$-equation~Eq.(\ref{b.eq}). 
For these reasons we limit ourselves to the minimum number of external sources, that is achieved in the Landau gauge.

We can now complete the set of functional identities defining the theory by extending the $X_{1,2}$-equations~\cite{Binosi:2022ycu}
in the presence of the additional external sources $N,P,L$, namely
\begin{itemize}
\item the $X_1$-equation of motion:
\begin{align}
    \frac{\delta \G}{\delta X_1}=
    \frac{1}{v} \square
    \frac{\delta \G}{\delta \bar c^*} 
    -
    \frac{1}{v} \square \Big ( L \frac{\delta \G}{\delta \phi^{\dagger *}} \Big )
    + 
    \square (P \omega) + \frac{1}{v} \square (N \phi^\dagger)  .    
    \label{X1.eq}
\end{align}
\item the $X_2$-equation of motion:
\begin{align}
    & \frac{\delta \G}{\delta X_2} =   \frac{1}{v} \square \frac{\delta \G}{\delta \bar c^*} -
    \frac{1}{v} \square \Big ( L \frac{\delta \G}{\delta \phi^{\dagger *}} \Big ) \nonumber \\
    & \qquad \quad 
 - \square X_1 - ( (1+z) \square + M^2) X_2 - v \bar c^* +
 \square (P \omega) + \frac{1}{v} \square (N \phi^\dagger) 
 .
	\label{X2.eq}
\end{align}

Notice that the $z$-term in Eq.(\ref{X2.eq}) is linear in $X_2$ and thus no new external source is needed in order to control its renormalization.
Notice that further bilinear terms $\sim X_2 \square^n X_2, n\geq 2$ could be added to the classical action while still modifying only the linear part in $X_2$ of the above equation. However, such higher-derivative contributions induce in the spectrum modes with negative metrics and lead to mathematical inconsistencies~\cite{Aglietti:2016pwz}.
\end{itemize}

\section{$z$-differential equation}\label{sec:zdiffeq}

The modified theory described in the above Section also obeys the 
$z$-differential equation presented in Ref.~\cite{Binosi:2022ycu}.
We give here a sketchy outline of the proof.

The second line of \1eq{tree.level} contains the deformation of the $X_2$-kinetic term controlled by the parameter $z$. When such term is switched off ($z=0$), we recover the power-counting renormalizable Higgs-Kibble model. 
At $z \neq 0$ the theory becomes non-renormalizable.
Nevertheless it can be uniquely defined by solving the $z$-differential equation presented in Ref.~\cite{Binosi:2022ycu}.
In particular, one defines the differential operator
\begin{align}
    {\cal D}_z^{M^2}=(1+z)\partial_z+M^2\partial_{M^2},
\end{align}
and notices that in the mass eigenstate basis 
the dependence on the parameter $z$ only arises via the $X_2$-propagator
\begin{align}
    \Delta_{X_2X_2}(k^2,M^2) = \frac{i}{(1+z)k^2-M^2}.
\end{align}
Then $\Delta_{X_2X_2}$ is an eigenvector of ${\cal D}_z^{M^2}$ with eigenvalue -1:
\begin{align}
    {\cal D}_z^{M^2}\Delta_{X_2X_2}(k^2,M^2)=-\Delta_{X_2X_2}(k^2,M^2).
\end{align}
Generalizing the argument to the decomposition in the number of the $\ell$-internal $X_2$-lines  of a one-particle-irreducible (1-PI) amplitude $\G^{(n)}_{\Phi_1\cdots\Phi_r}$ 
with external legs $\Phi_1(p_1),\dots,\Phi_r(p_r)$,
$\Phi_i$ denoting any of the fields and external sources of the theory with momenta $p_i$, $p_1 = -\sum_{i=2}^r p_i$, one gets
\begin{align}
    \G^{(n)}_{\Phi_1\cdots\Phi_r} = 
    \sum_{\ell\geq 0}\G^{(n;\ell)}_{\Phi_1\cdots \Phi_r} .
    \label{1pi.exp}
\end{align}
Then we have
\begin{align}
	{\cal D}^{M^2}_z \G^{(n;\ell)}_{\Phi_1\cdots\Phi_r} &= -
    \ell\G^{(n;\ell)}_{\Phi_1\cdots \Phi_r};& &\Longrightarrow& 
    {\cal D}^{M^2}_z \G^{(n)}_{\Phi_1\cdots\Phi_r} = -
    \sum_{\ell\geq 0}\ell\G^{(n;\ell)}_{\Phi_1\cdots \Phi_r}.
    \label{eq.1pi.exp}
\end{align}
The most general solution of this equation reads (indicating explicitly only the dependence on the parameters $z$ and $M^2$)
\begin{align}
	\G^{(n;\ell)}_{\Phi_1\cdots\Phi_r}(z,M^2)=\frac1{(1+z)^\ell}\G^{(n;\ell)}_{\Phi_1\cdots\Phi_r}(0,M^2/1+z).
	\label{genstr}
\end{align}
The above equation entails that the whole dependence of the amplitudes on $z$ is completely fixed by the 
amplitude in the r.h.s., evaluated at $z=0$ (power-counting renormalizable theory), after the replacement
$M^2 \rightarrow M^2/(1+z)$ and overall multiplication by the factor $1/(1+z)^\ell$.

The main advantage of the $X$-representation of the physical scalar mode is that the full dependence on the additional parameter $z$ is contained in the quadratic part of the classical vertex functional, so that in the perturbative expansion  the coupling $z$ will enter in the $X_2$-propagator but not in the  interaction vertices, contrary to what happens in the conventional $\phi$-representation of \1eq{d6.op}. 
This leads to the
derivation of Eq.(\ref{genstr}).

Moreover, since $X_2$ is gauge-invariant, one can prove that the Slavnov-Taylor (ST) identity in Eq.(\ref{sti}) can be graded
according to the number of $X_2$-lines, so that each
$\ell$-sector is separately invariant~\cite{Binosi:2022ycu}.

\section{Classical scalar potential and stability equation}\label{sec:potential}

We now consider the scalar potential in Eq.(\ref{tree.level}).
Let us denote by $\Sigma$ the gauge-invariant combination
\begin{align}
    \Sigma = \phi^\dagger \phi - \frac{v^2}{2} \, 
\end{align}
and set $X = X_1 + X_2$.
The classical scalar potential then reads
\begin{align}
V = \frac{M^2 -m^2}{2} X_2^2 + \frac{m^2}{2v^2} \Sigma^2 -
\frac{m^2}{v} X ( \Sigma - v X_2 ) \, .
\end{align}
The above quadratic form can be diagonalized via the redefinition
\begin{align}
    X_2 \rightarrow \tilde{X}_2 = X_2 + \frac{m^2}{M^2 - m^2} X \, , \qquad \Sigma \rightarrow \tilde{\Sigma} = \Sigma - v X
\end{align}
yielding
\begin{align}
    V = \frac{M^2- m^2}{2} \tilde{X}_2^2 + \frac{m^2}{2v^2}
    \tilde{\Sigma}^2 - \frac{1}{2} \frac{m^2 M^2}{M^2 - m^2}
    X^2 \, .
    \label{potential}
\end{align}
The minimum of the potential is reached when all fields
are equal to zero, namely $X_{1,2}=0$ and 
$$\tilde{\Sigma} = \phi^\dagger \phi - \frac{v^2}{2} = 0 \, ,$$
i.e. the usual condition on the v.e.v. of the Higgs field $\phi$.

In the set of extended field variables it turns out that
the potential $V$ is not positive definite, as can be seen
from Eq.(\ref{potential}).
Positivity of the term in $\tilde{X}_2$ implies 
$M^2> m^2$, so that the last term in Eq.(\ref{potential}) has a negative coefficient unless $m=0$.

In perturbation theory physical gauge-invariant observables are independent of $m$ (see Ref.~\cite{Binosi:2019nwz} for a detailed discussion).

The specific value of $m=0$ entails that the potential of the
theory in the extended formalism is positive-definite (under the assumption that $M>0$). 

It is interesting to notice that the vanishing of the potential along the $X_{1,2}$-directions is preserved also at the quantum level.
Indeed for $m=0$ by differentiating
Eq.(\ref{X1.eq}) w.r.t. $X_1(p_1), \dots, X_1(p_j), X_2(q_1), \dots, X_2(q_r)$ 
and by going on-shell at zero fields and external sources
one finds
($ p = -\sum_k p_k - \sum_l q_l $ by the conservation of the momentum)
\begin{align}
    & \frac{\delta^{(j+r+1)} \G^{(n)}}{\delta X_1(p_1) \dots \delta X_1(p_j) X_2(q_1)  \dots \delta X_2(q_r) \delta X_1(p)} \nonumber \\
    & \qquad \qquad \qquad = - p^2  \frac{\delta^{(j+r+1)} \G^{(n)}}{\delta X_1(p_1) \dots \delta X_1(p_j) X_2(q_1) \dots \delta X_2(q_r) \delta \bar c^*(p)} \, ,
\end{align}
so that on constant configurations at zero momentum the r.h.s. of the above equation vanishes. 
A similar argument applies to Eq.(\ref{X2.eq}), hence we get that also at the quantum level the potential
vanishes along the $X_{1,2}$-directions.

It is interesting to explore 
the limit $z\to \infty$ 
at fixed ratio $M^2/(1+z)$.
As a consequence of Eq.(\ref{genstr}),
all diagrams with at least one internal $X_2$-line vanish in this limit and one is left with a model with a physical scalar particle at
mass $M^2/(1+z)$ that does not propagate inside loops.
This is not the usual St\"uckelberg limit 
(where the mass of the Higgs field goes to infinity), since the physical scalar stays at a finite mass.
Yet it does not contribute to radiative corrections. 
In this sense it looks like a kind of  classical background field configuration.

\section{Conclusions}\label{sec:conclusions}

The extended formalism for the scalar sector has a number of specific features that are 
illustrated on the example of the Abelian HK model.
First of all, one obtains a novel differential equation connecting the power-counting renormalizable
theory at $z=0$ to the non power-counting renormalizable one at $z\neq 0$.
This is an important result that allows one to recover the infinite number of UV counter-terms, 
required to subtract the non power-counting renormalizable theory, in terms of the UV divergent amplitudes of the
power-counting renormalizable model by solving the $z$-differential equation
via its  most general solution
\begin{align}
   \G^{(n;\ell)}(z,M^2) = \frac{1}{(1+z)^\ell}\G^{(n;\ell)}\left(0,\frac{M^2}{1+z}\right),
\end{align}
namely by rescaling $M^2 \rightarrow M^2/(1+z)$ and by multiplying the $\ell$-amplitude 
at order $n$ in the loop expansion by the overall factor $1/(1+z)^\ell$.

Moreover, the ST identity holds true for each $\ell$-sector with a given number $\ell$ of internal $X_2$ lines.
Thus, even if power-counting renormalizabilty is lost, the model depends on the same number of physical parameters of its renormalizable realization at $z=0$ plus the $z$-parameter itself. 

Off-shell amplitudes also depend on a second (unphysical) mass parameter $m$. In the present paper we have
shown that for vanishing $m$ an additional stability equation exists for the scalar field $\phi$.
The stability of the classical potential along the flat $X_{1,2}$-directions for vanishing $m$ is preserved
at the quantum level, as a consequence of the $X_{1,2}$-equations.

The extension of this construction to non-Abelian gauge groups and in particular to the electroweak theory 
is currently under investigation.

\section*{Conflict of Interest}

The author declares that he has no conflicts of interest.


%

\end{document}